\documentclass[11pt,a4paper]{article}

\usepackage{graphicx}
\usepackage{hyperref}

\begin{document}
	
\title{Password Generators: Old Ideas and New\footnote{This
is the full version of a paper with the same title due to be
published in the proceedings of WISTP 2016 in September 2016.}}

\author{Fatma AL Maqbali and Chris J Mitchell\\
Information Security Group, Royal Holloway, University of
London\\
\href{mailto:me@chrismitchell.net}{\nolinkurl{me@chrismitchell.net}},
\href{mailto:fatmaa.soh@cas.edu.om}{\nolinkurl{fatmaa.soh@cas.edu.om}}
}
\date{15th July 2016}

\maketitle 	

\begin{abstract}
This paper considers password generators, i.e.\ systems
designed to generate site-specific passwords on demand.  Such
systems are an alternative to password managers.  Over the last
15 years a range of password generator systems have been
described.  This paper proposes the first general model for
such systems, and critically examines options for instantiating
this model; options considered include all those previously
proposed as part of existing schemes as well as certain novel
possibilities. The model enables a more objective and
high-level assessment of the design of such systems; it has
also been used to sketch a possible new scheme, AutoPass,
intended to incorporate the best features of the prior art
whilst also addressing many of the most serious shortcomings of
existing systems through the inclusion of novel features.
\end{abstract}

\section{Introduction}

Secret passwords remain a very widely used method for user
authentication, despite widely shared concerns about the level
of security they provide.  The remarkable persistence of
passwords has been discussed by Herley, van Oorschot and
Patrick \cite{Herley09}. There are many potentially replacement
technologies, including the use of some combination of
biometrics and trusted personal devices (e.g.\ as supported by
protocols such as FIDO UAF \cite{FIDOspec}), but it seems
likely that it will be some time, if ever, before passwords are
relegated to history.

Given their current wide use, and their likely continued use
for the foreseeable future, finding ways of improving the use
and management of passwords remains a vitally important issue
for real-world security.  In this paper we focus on an
important practical matter, namely how to make use of passwords
for authentication both more secure and more convenient. The
main focus is the use of passwords for remote user
authentication to a website. 	

Many technologies have been devised to either improve on the
level of security passwords provide, typically by: providing
alternative means of user authentication, through identity
management systems (such as OAuth \cite{RFC6749} or FIDO
\cite{FIDOspec}), or simply by making the use of strong, i.e.\
not readily guessable, passwords easier.  In this paper, in
line with the previous remarks about the continuing ubiquity of
passwords, we focus on the third of the above categories. In
particular, the fact that users are expected to memorise large
numbers of different strong passwords simply to go about their
daily business on the Internet, clearly forces users to
compromise their own security (around 10 years ago,
Flor\^{e}ncio and Herley, \cite{florencio07}, reported that
each user in a large scale study had about 25 accounts that
required passwords, and typed an average of 8 passwords per day
--- these numbers have probably risen significantly since their
study). The hope is that, in the short-to-medium term at least,
systems can be devised to make the user workload manageable
whilst still enabling the use of strong and diverse passwords. 	

One important category of schemes of this latter type are the
\emph{password managers} (what McCarney \cite{cite3} calls
\emph{retrieval password managers}).  A password manager stores
user passwords for a variety of sites, and produces them when
required (typically by auto-filling in login pages). Such
password managers fall into two main types, depending on
whether the passwords are stored locally or remotely on a
trusted server. Most browsers provide a local-storage password
manager as default functionality, and as a result local-storage
schemes are very widely used. 	

However, the shortcomings of password managers have also been
widely documented (see, for example, McCarney \cite{cite3}).
Passwords stored only on the user's platform restrict user
mobility, since they will not be available when a user user
switches, for example, from use of a laptop to a tablet or
phone.  However, if passwords are stored remotely `in the
cloud', then there is a danger of compromise through poorly
configured and managed servers.  Sadly there are real-world
examples of compromises of such password managers
\cite{Cluley13,Kelly13,Pauli15,Ragan15}.

Another, somewhat less well-studied, class of schemes, which
forms the main focus of this paper, involves generating
site-specific passwords on demand from a combination of inputs,
including those supplied by the user and those based on the
nature of the site itself. A number of individual schemes of
this type have been proposed but, apart from a brief summary by
McCarney \cite{cite3}, they have not been studied in a more
general setting. The main purposes of this paper are to (a)
provide a general model within which a range of (client side)
password generation schemes can be analysed, and (b) use this
model to propose a possible new system combining the best
features of existing schemes.  We believe that this is the
first time this class of schemes has been considered in a
unified way, and by doing so we gain new insights into their
design and implementation.

The rest of the paper is organised as follows.
Section~\ref{model} introduces a general model for
\emph{password generators}; key examples of such schemes are
also given. This is followed in section~\ref{components} by a
review of the options for the chief components of the model,
again referring to existing example schemes.
Section~\ref{assessment} then addresses the advantages and
disadvantages of these options. In section~\ref{improving} we
consider enhancements to the operation of the system which
mitigate some of the identified disadvantages.  The lessons
from our assessments, together with these novel enhancements,
are incorporated into a high-level design for a novel scheme,
AutoPass, described in section~\ref{novel}.
Section~\ref{conclusion} concludes the paper and provides
directions for further research.
 	
\section{Password Generators --- A General Model} \label{model}

This paper is concerned with \emph{password generators}, i.e.\
schemes designed to simplify password management for end users
by generating site-specific passwords on demand from a small
set of readily-memorable inputs.  Note that the term has also
been used to describe schemes for generating random or
pseudorandom passwords which the user is then expected to
remember; however, we use the term to describe a system
intended to be used whenever a user logs in and that can
generate the necessary passwords on demand and in a repeatable
way.

A variety of such schemes have been proposed in recent years
--- however, this paper focusses on the general
properties of such schemes, and the various options for their
operation. We start by presenting a general model for such
schemes, which we then use as a framework for analysing
possible scheme components. We observe that the general class
of such schemes has been briefly considered previously by
McCarney \cite{cite3} under the name \emph{generative password
managers}.

\subsection{A Model}

A password generator is functionality implemented on an
end-user platform to support password-based user authentication
to a remote server (assumed to be a web site, although most of
the discussion applies more generally). This functionality
generates, on demand, a site-unique password for use in
authentication. Clearly this password also needs to be
available to the web site authenticating the user --- the
nature of the \emph{registration} step, in which the password
is set up, is discussed further in section~\ref{registration}
below.  A password generator has the following components.

\begin{itemize}
\item A set of \emph{input values} is used to determine the
    password for a particular site.  Some values must be
    site-specific so that the generated password is
    site-specific.  The values could be stored (locally or
    online), based on characteristics of the authenticating
    site, or user-entered when required. Systems can, and
    often do, combine these types of input.
\item A \emph{password generation function} combines
		the input values to generate an
		appropriate password.  This function could operate in a
		range of ways depending on the requirements of the web
		site doing the authentication. For
		example, one web site might forbid the inclusion of
		non-alphanumeric characters in a password, whereas
		another might insist that a password contains
		at least one such character.  To be broadly
		applicable, a password generation function must
		therefore be customisable.
\item A \emph{password output method} enables the generated
    password to be transferred to the authenticating site.
    This could, for example, involve displaying the
    generated password to the user, who must then type (or
    copy and paste) it into the appropriate place.
\end{itemize}

All this functionality needs to be implemented on the user
platform. There are various possibilities for such an
implementation, including as a stand-alone application or as a
browser plug-in. Each of these aspects of the operation of a
password generator are discussed in greater detail in
section~\ref{components} below.

\subsection{Examples}

Before proceeding we briefly outline some existing proposals
for password generation schemes conforming to the above model.
The schemes are presented in chronological order of
publication.  Note that the functional components of the
various examples are considered in greater detail in
section~\ref{components} below. 	
\begin{itemize}
\item The \emph{Site-Specific Passwords (SSP)} scheme
    proposed by Karp \cite{cite4} in 2002/03 is one of the
    earliest proposed schemes of this general type.  SSP
    generates a site-specific password by combining a
    long-term user master password and an easy-to-remember
    name for the web site, as chosen by the user.
\item \emph{PwdHash}, due to Ross et al.\ \cite{cite5},
    generates a site-specific password by combining a
    long-term user master password, data associated with
    the web site, and (optionally) a second global password
    stored on the platform.
\item The 2005 \emph{Password Multiplier} scheme of
    Halderman, Waters and Felten, \cite{cite6}, computes a
    site-specific password as a function of a long-term
    user master password, the web site name, and the user
    name for the user at the web site concerned.
\item Wolf and Schneider's 2006 \emph{PasswordSitter}
    \cite{Wolf06} scheme generates a site-specific password
    as a function of a long-term user master password, the
    user identity, the application/service name, and some
    configurable parameters.
\item \emph{Passpet}, due to Yee and Sitaker \cite{cite7}
    and also published in 2006, takes a very similar
    approach to SSP, i.e.\ the site-specific password is a
    function of a long-term user master password and a
    user-chosen name for the web site known as a
    \emph{petname}.  Each petname has an associated icon,
    which is automatically displayed to the user and is
    intended to reduce the risk of phishing attacks.
\item \emph{ObPwd}, due to Mannan et al.
    \cite{Biddle11,cite2,Mannan08,cite1}, first surfaced in
    2008.  It takes a somewhat different approach by
    generating a site-specific password as a function of a
    user-selected (site-specific) object (e.g.\ a file),
    together with a number of optional parameters,
    including a long-term user password (referred to as a
    \emph{salt}), and the web site URL.
\item Finally, \emph{PALPAS} was published in 2015
    \cite{horsch15}. PALPAS generates passwords complying
    with site-specific requirements using server-provided
    password policy data, a stored secret master password
    (the \emph{seed}), and a site- and user-specific secret
    value (the \emph{salt}) that is synchronised across all
    the user devices using the server.
\end{itemize}

\subsection{Registration and Configuration}
\label{registration} \label{configuration}

In this paper we only consider schemes whose operation is
completely transparent to the website which is authenticating
the user. As a result, the `normal' website registration
procedure, in which the user selects a password and sends it to
the site, is assumed to be used.  This, in turn, typically
means that the password generation process needs to be in place
\emph{before} the registration procedure, or at least that
introduction of the password generator requires the user to
modify their account password.  This potential disadvantage of
password generators, together with possible ways of avoiding
the need to change passwords, is examined in
sections~\ref{other-issues} and~\ref{novel-types} below.

There is a potential need for a password generator to store
configuration data. Such data can be divided into two main
types:
\begin{itemize}
\item \emph{global configuration data}, i.e.\ values unique
    to the user and which are used to help generate all
    passwords for that user, and
\item \emph{site-specific configuration data}, i.e.\ values
    used to help generate a password for a single website,
    which are typically the same for all users.
\end{itemize}
Not all schemes use configuration data, although producing a
workable system without at least some global configuration data
seems challenging. However, the use of configuration data is
clearly a major barrier to portability. That is, for a user
employing multiple platforms, the configuration data must be
kept synchronised across all these platforms, a non-trivial
task --- exactly the issue addressed in a recent paper by
Horsch, H\"{u}lsing and Buchmann \cite{horsch15}.

\section{Components of the Model}  \label{components} 	

We next consider in greater detail options for the components
of the model.

\subsection{Inputs to Password Generation}  \label{inputs}

We first consider the input values to the password generation
process.  The following data input types have been employed in
existing schemes.

\begin{itemize}
\item A \textbf{master password} is a user-specific
    long-term secret value.  This could either be a
    \textbf{user-entered password}, i.e.\ entered by the
    user whenever a password is to be generated, or a
    \textbf{stored password}, i.e.\ a user-specific secret
    value stored as global configuration data.  Note that
    this could be augmented by use of a (not necessarily
    secret) \textbf{user constant}, i.e.\ a further global
    configuration value entered by the user and which
    ensures that the passwords generated by this instance
    of the scheme are different to those generated by
    another instance, even if the same password is used.

\item A \textbf{site name} is a name for the site using the
    password for user authentication. This could take a
    variety of forms, including a \textbf{user site name},
    i.e.,\ a name for a site chosen by a user, all or part
    of the site's \textbf{URL}, or a \textbf{site-specific
    secret}, e.g.\ a random value associated with the site
    URL.

\item A \textbf{digital object} is anything available on
    the user platform which could be used as input to the
    password generation process, e.g.\ a file or a selected
    block of text on the target web site. Typically the
    user would be expected to use a different object (or
    set of objects) for each web-site.

\item A \textbf{password policy} is information governing
    the nature of the password generated, e.g.\ the set of
    acceptable symbols.


\end{itemize}	

Table~\ref{table:inputs} summarises the sets of data inputs
used by existing password generator schemes.  Items given in
square brackets ([thus]) are optional inputs.

\begin{table}[hbtp]
\caption{Inputs to password generation process}
\label{table:inputs}
\centering
\begin{tabular}{|l|l|}

\hline

\textbf{Scheme} & \textbf{Input data} \\ \hline

ObPwd, \cite{cite2} &
\parbox{6cm}{{digital object}, [{URL}]} \\ \hline

PassPet, \cite{cite7}  & \parbox{6cm}{user constant, user-entered password, user site name} \\
\hline

Password Multiplier, \cite{cite6} & user-entered
password, user site name \\ \hline

PasswordSitter, \cite{Wolf06} & user-entered password,
user site name \\ \hline

PwdHash, \cite{cite5} & \parbox{6cm}{user-entered password,
URL } \\
\hline

Site-Specific Passwords, \cite{cite4} & user-entered password,
user site name \\ \hline

PALPAS, \cite{horsch15} & \parbox{6cm}{stored password,
site-specific secret, password policy} \\
\hline
		
\end{tabular}
\end{table}

Two other possible types of value that could be used to help
generate a site-specific password, and that have not previously
been proposed, are as follows.
\begin{itemize}
\item The first is to employ the \textbf{user name} for the
    user at the site concerned.  Typically this will be
    something readily memorable by the user, but will often
    vary from site to site.
\item A second rather more complex option would be to
    provide a local front end for a graphical password
    system (see, for example, \cite{dirik07,takada03}), and
    to use this to generate a bit-string to be input to
    password generation.  This possibility is not explored
    further here.
\end{itemize}

\subsection{Generating the Password}  \label{generation}

A number of approaches can be used to combine the various
inputs to generate a password.  They all involve a two-stage
process, i.e.\ first combining the inputs to generate a
bit-string, and then formatting and processing the bit-string
to obtain a password in the desired format.  Typically a target
format is a string of symbols of a certain length, where each
symbol is, for example, numerals only, alphanumeric characters,
or alphanumerics together with punctuation.

Possibilities for the first stage include the following.
\begin{itemize}
\item \textbf{one-level-hash}, i.e.\ concatenating the
    various inputs and applying a cryptographic
    hash-function.  An alternative of this one-level type
    would be to use part of the input as a key, and to then
    generate an encryption or MAC on the remainder of the
    input. Examples of this general approach include
    PwdHash\cite{cite5}, ObPwd \cite{cite2} and SSP
    \cite{cite4}.  PasswordSitter, \cite{Wolf06}, uses AES
    encryption as an alternative to a hash-function, where
    the AES key is derived from the master password.
\item A widely discussed alternative is the
    \textbf{two-level-hash}. In this case, one or more of
    the inputs are concatenated and input to a
    cryptographic hash-function which is then iterated some
    significant number of times.  The output of this
    multiple iteration is then concatenated with the other
    inputs and hashed to give the output.  This two-level
    multiple iteration process is designed to slow down
    brute force attacks.  Examples of systems adopting such
    an approach include PassPet \cite{cite7} and Password
    Multiplier \cite{cite6}.
\end{itemize}

The main approach to the second stage employs some form of
\textbf{encoding}, in which the output from the first stage is
formatted to obtain the desired password from a site-specific
character set, perhaps also satisfying certain rules (e.g.\
mandating the inclusion of certain classes of character). Some
existing schemes, such as ObPwd [ref], allow this to be
parameterised in order to meet specific website requirements.
Horsch et al. \cite{horsch16} go one step further and propose
an XML syntax, the \emph{Password Requirements Markup Language
(PRML)}, designed specifically to enable such requirements on
passwords to be formally specified. Such password policy
statements constitute site-specific configuration data.

\subsection{Password Output and Use} \label{output}

There are a number of ways in which a generated password could
be transferred to the password field of a website login page.

\begin{itemize}
\item The simplest is \textbf{manual copy and paste}, where
    the password generation software displays the generated
    password to the user who manually copies it into the
    login page. This approach is used by SSP \cite{cite4}.
\item A slightly more automated approach is \textbf{copy to
    clipboard}, in which the generated password is copied
    into the clipboard for future use. For security reasons
    the password can be made to only reside in the
    clipboard for a limited period, e.g.\ in PasswordSitter
    the generated password is saved to the clipboard for 60
    seconds before being deleted \cite{Wolf06}.
\item The simplest approach for the user is probably
    \textbf{automatic copying to the target password
    field}.  This can either be done automatically with no
    user intervention, as is the case for PwdHash in the
    web page implementation \cite{cite5} and the ObPwd
    Firefox browser extension \cite{cite2}.  Alternatively
    it can require the user to perform an action such as
    clicking a specific key combination before copying; for
    example, PassPet requires the user to click on a screen
    button, \cite{cite7}, and Password Multiplier,
    \cite{cite6}, requires the user to double click the
    password field or press \textit{ctrl+P} to trigger
    password copying.
\end{itemize}

\subsection{Approaches to Implementation}
\label{implementation}

Password generation software can be implemented in a range of
ways.

\begin{itemize}
\item There are a number of advantages to be derived from
    implementing the password generator as a
    \textbf{browser add-on}, e.g.\ as a \textbf{browser
    plug-in}, \textbf{browser extension} or \textbf{signed
    browser applet}. Many existing password generator
    schemes adopt this approach, at least as one option,
    including \cite{cite6,cite5,Wolf06,cite7}.

\item An alternative is to implement the scheme as a
    \textbf{stand-alone application}, e.g.\ to run on a
    phone, tablet or desktop. The user would need to
    install the application.  Such an approach is adopted
    by SSP; ObPwd, \cite{cite2}, is also available as both
    a browser extension and a mobile app.

\item A somewhat different approach is to implement the
    scheme as a \textbf{web-based application}, either
    running on a remote server or executing on the user
    platform in the form of dynamically downloaded
    JavaScript.
\end{itemize}

\section{Assessing the Options}  \label{assessment}

For each of the main components we assess the main advantages
and disadvantages of the possibilities described in
section~\ref{components}.

\subsection{Inputs}

Three main options for the inputs to the password generation
process were described in section~\ref{inputs}.  We consider
them in turn.

\begin{itemize}
\item The use of a \textbf{master password} of some kind
    seems highly advantageous; the main issue is how it is
    made available when required, i.e.\ whether to store it
    long-term in the software as global configuration data
    or to employ a user-entered value. Both possibilities
    have pros and cons. Long-term storage maximises user
    convenience, but the secret is now at greater risk of
    exposure and the system is now inherently less
    portable, not least because the user may well forget
    the value if it does not need to be entered regularly.
    Conversely, user entry reduces convenience but possibly
    improves security, although the entry process itself is
    now prone to eavesdropping, either visually or using
    some kind of key-logger. Perhaps the best possibility
    might be to combine the two, i.e.\ to use two secrets
    --- one stored long-term on every device employed by
    the user and the other entered by the user whenever the
    system is activated.  Such an approach is implemented
    in the PALPAS system, \cite{horsch15}, where a
    user-entered password is employed to generate a key for
    encrypting and decrypting the locally stored master
    secret (or \emph{seed}).

    The addition of some kind of \textbf{user constant},
    i.e.\ a not necessarily secret user-specific value,
    also seems reasonable.  The obvious location for this
    is as global configuration data, although this again
    may have a portability impact.

\item The inclusion of a \textbf{site name} was the second
    option considered in section~\ref{inputs}. The use of
    such an input is highly desirable since it will make
    generated passwords site-specific. The site name could
    be a \textbf{user site name}, i.e.\ a name for the site
    selected by the user, the site's \textbf{URL}, or a
    value indexed by the URL\@. One disadvantage of a
    \textbf{user site name} is that it must be remembered
    (and entered) by the user. Use of the \textbf{URL}
    avoids the latter problem by potentially being
    available automatically to the password generator.  It
    also prevents phishing attacks in which a fake site
    attempts to capture user credentials for the site it is
    imitating, since the URL of the phishing site, and
    hence the generated password, will be different to that
    of the genuine site, \cite{cite5}. However, use of the
    site URL also has issues, \cite{cite4}, since the URL
    of a site can change without notice, meaning the
    generated password would also change and the system
    would fail. This latter issue can be at least partially
    addressed by using only the first part of the URL.

\item The third possibility is use of a \textbf{digital
    object}.  Perhaps its main advantage is that it
    potentially introduces a major source of entropy into
    password generation \cite{cite2,cite1}. Such an input
    also offers a way of making passwords site-specific,
    although it requires users to choose a different object
    for every site (and it is not clear that all users will
    do so). However, there are two major disadvantages with
    such an approach.  Firstly, the object used must always
    be accessible, significantly restricting user choice
    especially on small platforms such as phones. Secondly,
    the user must remember which object is used with which
    site, a task which users may find as hard as
    remembering site-unique passwords (depending on the
    psychology of the individuals involved).
\end{itemize}

\subsection{Generating the Password}

As discussed in section~\ref{generation}, password generation
is typically a two-stage process: first combine the inputs to
generate a bit-string, and second use the bit-string to
generate a password.  The first stage involves either a single
level or a two-level hash.  The two-level approach has the
advantage of offering a limited degree of protection against
brute-forcing of a master secret, \cite{cite10,cite6,cite7}.
The only disadvantage is a slight delay in the password
generation process itself, but this can be made small enough to
be barely noticeable.

The second \textbf{encoding} step is more problematic. Websites
have widely differing password requirements and rules. The
encoding scheme must generate passwords tailored to
site-specific requirements (referred to by Horsch, H\"{u}lsing
and Buchmann \cite{horsch15} as the \emph{password policy}).
This in turn typically requires the password generator to store
site-specific password policies as site-specific configuration
data, potentially reducing the portability of the password
generator since each instance must be locally configured.

\subsection{User Interface Operation}

Section~\ref{output} describes a range of possible approaches
for transferring a generated password to the password field in
the website login page.

\begin{itemize}
\item The \textbf{manual copy and paste} approach is
    clearly the simplest to implement.  However, apart from
    being the least user-friendly, it has a security
    deficiencies. Most seriously, a user could be tricked
    by a fake page to reveal their password for a genuine
    site.  There is also a serious possibility of
    eavesdropping (or `shoulder-surfing').

\item The \textbf{copy to clipboard} technique is much more
    convenient for the user, but is still prone to fake
    website attacks.  Also, an attacker might be able to
    launch an attack to learn the clipboard contents. It
    would appear to be good practice to restrict the period
    of time during which the generated password is in the
    clipboard, as implemented in PasswordSitter
    \cite{Wolf06}.

\item The most convenient approach for the user is
    undoubtedly \textbf{automatic copying to the target
    password field}.  If the password generator is aware of
    the URL of the web page (which is likely if it can
    access the page to autofill the password) then this can
    mitigate fake website attacks. The major disadvantage
    relates to implementability --- practicality depends
    very much on how the password generator is implemented
    (as discussed in
    section~\ref{implementation-discussion} below).  Such a
    solution might require the user to perform a specific
    action to trigger automatic copying of the password;
    this would have the advantage of giving the user some
    control over the process.
\end{itemize}

\subsection{Implementation}  \label{implementation-discussion}

We conclude this assessment of the various password generator
options by considering the ways in which it might be
implemented.

\begin{itemize}

\item The \textbf{browser add-on} approach has been widely
    advocated in the literature, and has a number of
    advantages.  When implemented in this way, a password
    generator can readily automate key tasks, including
    detecting the password field, discovering the site URL,
    and filling in the password field automatically.
    However, the use of multiple browsers across multiple
    platforms may cause incompatibility problems, and
    require the development and use of multiple instances
    of the scheme.  Also, the lack of an easily accessible
    user interface may also make configuration difficult
    for non-expert users.

\item A \textbf{stand-alone application} is the obvious
    alternative to a browser add-on. One advantage of such
    an approach compared to a browser add-on relates to the
    user interface; a stand-alone application is likely to
    have a richer user interface, easing its use and
    configuration. However, such an application may be
    unable to automatically perform some of the tasks which
    can readily be performed by a browser add-on, such as
    automated password field detection and password input.
    Stand-alone applications will also have portability
    issues, with a different application needed for each
    platform type.

 \item The third possibility is a \textbf{web-based
     application}.  Such an approach has the great
     advantage of seamless portability --- it would be
     instantly usable on any platform at any time.  Just
     like a browser add-on, it could also enable automation
     of key tasks such as URL detection and automatic
     password completion, \cite{cite6}.  However, there are
     also serious disadvantages.    A usability study
     conducted by Chiasson et al., \cite{cite10}, revealed
     that users had difficulty in locating the PwdHah
     website.  Most seriously, the web implementation will
     potentially have access to all the user's passwords,
     as well as the values used to generate them.  Of
     course, an application could be implemented, e.g.\
     using JavaScript, to perform all the necessary
     calculations on the user machine, and not on a remote
     server --- however, the capability would remain for
     the website to transparently eavesdrop on the process
     whenever it wished.

\end{itemize}

\subsection{Other Issues}  \label{other-issues}

Before proceeding we mention certain other usability and
security issues which can arise, potentially regardless of the
options chosen.

\begin{description}

\item[Setting and updating passwords]  If a user is already
    using the password generator when newly registering
    with a website, there is clearly no problem --- the
    user can simply register whatever value the system
    generates.  However, if the user has selected and
    registered passwords with a range of websites
    \emph{before} starting use of the password generator,
    then all these passwords will need to be changed to
    whatever the password generator outputs
    --- this could be highly inconvenient if a user has
    established relationships with many sites, and could
    present a formidable barrier to adoption of the system.

    Somewhat analogous problems arise if a user decides to
    change a website password, e.g.\ because the site
    enforces periodic password changes.  The only
    possibility for the user will be to change one of the
    inputs used to generate the password, e.g.\ the object
    (if a digital object is used as an input) or a user
    site name.  Password change could even be impossible if
    the user does not choose any of the site-specific
    inputs used to generate a password.

\item[Using multiple platforms] In various places we have
    noted issues that can arise if a user employs multiple
    platforms.  This is a general problem that arises in
    particular if any kind of configuration data is used by
    the system.  We discuss possible solutions to this
    problem in section~\ref{server} below.

\item[Password policy issues] A further general problem
    relates to the need to generate passwords in a
    site-specific form.  This password policy issue has
    been explored extensively by Horsch and his co-authors
    \cite{horsch15,horsch16}.  We return to this issue in
    section~\ref{novel-types} below.

\end{description}

\section{Improving System Operation}  \label{improving}

We next consider a range of ways of addressing some of the
problems we have identified.  Some of these techniques have
already been proposed, although not precisely as we describe
them here. In section~\ref{novel} we consider how these
measures might be integrated into a novel system.

\subsection{Novel Types of Configuration Data}
\label{novel-types}

We have already mentioned certain types of configuration data,
including global data, such as master secrets, and
site-specific data, such as password policy values (possibly
specified in PRML \cite{horsch16}).  We now introduce two new
configuration data types, whose use can address some of the
identified issues.

\begin{itemize}

\item A \emph{password offset}, a type of site-specific
    configuration data, can address the issue that a user
    may already have a well-established set of passwords
    which he/she does not wish to change; also, in some
    cases specific password values may be imposed on users.
    Additionally, users may wish, or be required to, change
    their passwords from time to time. As we have already
    observed, addressing such requirements with a password
    generation scheme is problematic.

    The idea of a password offset is as an input to the
    second stage of password generation.  The first stage
    generates a bit-string, and the second stage converts
    this to a password with specific properties, as
    specified by the password policy. The password offset
    induces the second stage to generate a specific
    password value.  For example, suppose that a password
    policy dictates that a password must be a string of
    lower and upper case letters and numerals, and suppose
    each such character is internally represented as a
    numeric value in the range 0--61. After converting the
    bit-string to a string of alphanumeric characters of
    the desired length, and given a `desired password'
    consisting of an alphanumeric string of the same
    length, the password offset could simply be the
    character-wise modulo 62 difference between the two
    strings\footnote{Such an idea ia widely implemented to
    enable credit/debit card holders to select their own
    PIN value.}.  Changing a password can now be readily
    implemented by changing the offset, either to a random
    value (thereby randomising the password choice), or to
    a chosen value (if the new password value is to be
    fixed by the user).

    If implemented appropriately, this offset is not hugely
    sensitive, since it need not reveal anything about the
    actual password value.  Of course, if an `old' password
    is compromised, and the old and new offsets are also
    revealed, then this could compromise the new password
    value.

\item It is also possible to envisage a scheme where a
    password for one site is generated using a different
    set of input types to those used to generate a password
    for another site.  For example, a password for a
    particularly mission-critical site (e.g.\ a bank
    account) might be generated using a large set of input
    values, e.g.\ including a digital object, whereas a
    password for a less sensitive site could be generated
    using a master secret and site name only.  Such a
    possibility could readily be captured using
    site-specific configuration data which we refer to as
    \emph{password input parameters}.

\item A system might also store \emph{password reminders}
    as site-specific configuration data.  For example, when
    choosing a digital object to generate a password, the
    user could be invited to specify a word or phrase to
    act as a reminder of the chosen value (without
    specifying it precisely).  This could then be revealed
    on demand via the password generator user interface.

\end{itemize}

\subsection{Use of a Server}  \label{server}

We have already observed that storing configuration data on a
user platform creates a major barrier to portability.  It also
poses a certain security risk through possible platform
compromise, although, apart from the master secret, much of the
configuration data we have discussed is not necessarily
confidential.

The `obvious' solution to this problem is to employ a server to
store configuration data, or at least the less sensitive types
of configuration information, much as many password managers
keep user passwords in the cloud. That is, while it would seem
prudent to at least keep a master secret on the user platform,
all the site-specific configuration data could be held in the
cloud. This type of solution is advocated by Horsch and his
co-authors \cite{horsch15,horsch16}.

If the scope of the site-specific configuration data can be
kept to non-confidential values, then there is no need for a
highly trusted server, a great advantage by comparison with
some of the server-based password managers.  Also, use of a
server need not significantly impact on password availability,
since a password generator application could cache a copy of
the configuration data downloaded from the server, addressing
short term loss of server availability.  Loss of network
availability would not be an issue, since in such a case remote
logins would in any event not be possible, i.e.\ passwords
would not be needed.

\section{AutoPass: A New Proposal}  \label{novel}


We now describe the design of AutoPass (from `automatic
password generator'), a novel password generation scheme
combining the best features of the prior art together with the
novel ideas introduced in this paper, particularly those
devised to address some of the shortcomings of previously
proposed schemes.  Full specification and implementation of
AutoPass remains as future work.

AutoPass uses all the types of input given in
section~\ref{inputs} to generate a password, since they all
contribute to security in different ways. Following the
approach of PALPAS, \cite{horsch15}, we also propose to make
use of a server to store non-sensitive configuration data, such
as website password policies.

\subsection{Operation}

Following the model of section~\ref{model}, in order to
describe AutoPass we need to define: (a) the input types, (b)
how the password is generated, and (c) how the password is
output, together with the implementation strategy.  We cover
each of these points in turn.  Since we also propose the use of
a cloud service to support AutoPass operation, we also briefly
sketch the operation of this service.

\begin{itemize}
\item As far as the \textbf{inputs} are concerned, we
    propose the use of a \textbf{master password}, where a
    password is stored by the system (as global
    configuration data) and a password (or PIN) must also
    be entered by the user.  We further propose the use of
    the first part of the \textbf{URL} for the site, where,
    depending on the implementation, this should also be
    stored as part of the site-specific configuration and
    used to retrieve the other site-specific data. The
    master password can be held encrypted by a key derived
    from the user password. We also propose the optional
    use of a \textbf{digital object}, where whether or not
    the option is used for this site is indicated in the
    site-specific configuration data.
\item The first stage of \textbf{password generation}
    adopts the two-level hash approach, giving some
    protection for the master secret against brute force
    attacks.  The second stage, i.e. \textbf{encoding},
    uses the AutoPass cloud service to retrieve information
    about the password policy of the web site being visited
    (analogously to PALPAS \cite{horsch15}). This password
    policy could, for example, be encoded using PRML,
    \cite{horsch16}.  It also uses other cloud-stored
    configuration data, notably the password offset,
    password input parameters, and password reminders
    introduced in section~\ref{configuration}.
\item The precise option for \textbf{password output and
    use} depends on the implementation.  Where possible,
    autofilling the password is desirable; where this is
    impossible, the copy to clipboard/paste buffer approach
    is advocated.
\item \textbf{Implementation} as a browser add-on is
    probably the best option, not least in giving simple
    access to the web page of the target site, although a
    range of options may need to be pursued depending on
    the platform type.
\end{itemize}

We next consider the operation of the AutoPass Cloud Service.
It will be required to store two main types of data:
\begin{itemize}
\item \emph{User-independent data} will be accessed by all
    users of AutoPass, and will include site-specific data
    such as password policies.  This data is completely
    non-sensitive.  Even if it could be corrupted by a
    malicious party, it would at worst cause a denial of
    service.
\item \emph{User-specific data} will only be accessed by a
    single user, and will include a range of password
    configuration data.  Although this data is not highly
    confidential, access to it will need to be restricted
    to the user to whom it belongs, e.g.\ via a one-off
    login process via the local AutoPass application (with
    access permissions encoded in a cookie stored in the
    user platform).
\end{itemize}

Any use of a cloud service has associated risks arising from
non-availability of the service.  However, this can largely be
addressed through caching.  The local AutoPass app should
maintain a copy of the data downloaded from the cloud service
--- since this data is not likely to change very quickly, the
cached data should normally be sufficient to maintain normal
operation.

To avoid risks arising from fake AutoPass services, e.g.\ using
DNS spoofing, the cloud service could sign all the data it
provides, and the AutoPass app could verify signatures using a
built-in copy of the cloud service public key.  However,
whether this is practically viable or necessary is a topic for
further study.

\subsection{Assessment}

The AutoPass system has been designed to incorporate both the
best features of the existing password generation schemes and
certain novel features, notably the use of password
configuration data (see section~\ref{improving}).  Of course, a
full assessment of AutoPass will require the design and
implementation of at least one prototype, and subsequent user
testing. Nonetheless, we can at least reconsider the issues we
identified in existing systems (see section~\ref{assessment})
and analyse to what extent AutoPass addresses these concerns.

By using a combination of stored secret and memorised
password/PIN as inputs to the generation process, we enable
strong passwords to be generated while protecting against
compromise through platform loss.  Use of the URL enables
automatic generation of site-specific passwords, and the
optional use of digital objects enables passwords of varying
strength and diversity to be generated without imposing an
unnecessarily onerous load on the user for `everyday' use.  Use
of the URL has residual problems, notably if a website URL
changes, but user site names also have problems. In this
connection, an advantage of the AutoPass approach is that
password offsets enable the generated password to remain
constant even if the URL changes.

The use of cloud-served password policies solves problems
relating to site-specific requirements for passwords.  The
problems arising from a desire to continue using existing
passwords and the potential need to change passwords are
conveniently addressed by the use of cloud-stored password
configuration data.  Password generation/synchronisation issues
arising from the use of multiple platforms can also be
addressed through the use of a cloud service.  Of course, use
of a cloud service brings with it certain availability and
security risks.  However, by only storing non-sensitive data in
the cloud, and also using caching, these problems are largely
addressed.

\section{Concluding Remarks}  \label{conclusion}

We introduced a general model for password generation, and
considered all the existing proposals in the context of this
model.  The model enables us to analyse the advantages and
disadvantages of a range of approaches to building such
systems.  It also enables us to propose certain new options to
enhance such schemes.

This paper is primarily theoretical; it provides a general
framework to study the design of password generation schemes.
The operation of a novel scheme, AutoPass, has been sketched,
but has not yet been tested in practice. The obvious next step
is to prototype aspects of AutoPass, both to verify that the
underlying ideas work and also to provide a basis for practical
user trials.

\bibliographystyle{plain}
\bibliography{mybib-v3-4}

\end{document}